\documentclass[conference]{IEEEtran}
\IEEEoverridecommandlockouts

\usepackage{comment}
\usepackage[super]{nth}
\usepackage{cite}
\usepackage{amsmath,amssymb,amsfonts}
\usepackage{algorithmic}
\usepackage{graphicx}
\usepackage{textcomp}
\usepackage{xcolor}
\def\BibTeX{{\rm B\kern-.05em{\sc i\kern-.025em b}\kern-.08em
    T\kern-.1667em\lower.7ex\hbox{E}\kern-.125emX}}
\begin{document}

\title{Hybrid Artifact Detection System for Minute Resolution Blood Pressure Signals from ICU\\

\thanks{This research was funded in part by an EPSRC Doctoral Training Partnership PhD studentship and The Carnegie Trust for the Universities of Scotland (RIG009251). For the purpose of open access, the author has applied a Creative Commons Attribution (CC BY) licence to any Author Accepted Manuscript version arising from this submission.}
}

\author{\IEEEauthorblockN{Hollan Haule\IEEEauthorrefmark{1}, Evangelos Kafantaris\IEEEauthorrefmark{1}, Tsz-Yan Milly Lo\IEEEauthorrefmark{2}, Chen Qin\IEEEauthorrefmark{1}, and Javier Escudero\IEEEauthorrefmark{1}}
\IEEEauthorblockA{\IEEEauthorrefmark{1}\textit{School of Engineering, IDCOM, University of Edinburgh, Edinburgh, UK}}

\IEEEauthorblockA{\IEEEauthorrefmark{2}\textit{Centre of Medical Informatics, Usher Institute, University of Edinburgh, Edinburgh, UK}}}

\maketitle

\begin{abstract}
    Physiological monitoring in intensive care units (ICU) generates data that can be used in clinical research. However, the recording conditions in clinical settings limit the automated extraction of relevant information from physiological signals due to noise and artifacts. Therefore, removing artifacts before clinical research is essential. Manual annotation by experienced researchers, which is the gold standard for removing artifacts, is time-consuming and costly due to the volume of the data generated in the ICU. In this study, we propose a hybrid artifact detection system that combines a Variational Autoencoder with a statistical detection component for the labelling of artifactual samples to automate the costly process of cleaning physiological recordings. The system is applied to minute-by-minute mean blood pressure signals from an intensive care unit dataset. Its performance is verified by manual annotations made by an expert. We benchmark the performance of our system with two other systems that combine an ARIMA or an autoencoder based model with our statistical detection component. Our results indicate that the system consistently achieves sensitivity and specificity levels over 90\%. Thus, it provides an initial foundation to automate data cleaning in recordings from ICU.
\end{abstract}


 \section{Introduction}
 Bedside physiological monitoring of patients in Intensive Care Units (ICU) generates wealth of data. The effective use of such data can lead to clinical research towards creating early and personalized disease interventions. In the case of Traumatic Brain Injury (TBI), real-time monitoring is crucial for early detection and prevention of secondary insults \cite{Jones2003a}. TBI accounts most injury-related deaths in Europe \cite{Tagliaferri2006}. Hence, it is essential to develop procedures to improve the clinical management of TBI patients.
 
 Typically, ICU patient monitoring involves collection of continuous recording of patients' physiological state signals such as blood pressure, core temperature, and blood gas levels \cite{Sorani2007}. However, the intrinsic conditions of clinical work in the ICU lead to low-data quality. Position changes, clinical interventions, probe change, and accidental probe dislodgement lead to  artefacts in data \cite{Johnson2016}. Procedures such as arterial blood sampling may also temporarily affect the reliability of recordings. 
 
 
 Senior clinicians are trained to ignore artifacts in clinical decision-making. However, false alarms triggered by artifacts may lead to ``alarm fatigue'' \cite{Johnson2016}, which risks patient outcomes. Moreover, any utilization of raw ICU data for offline clinical research is also limited due to the low-data quality. Currently, the ``gold standard'' for removing artifacts is manual annotation by experienced researchers, which is time-consuming \cite{Alkhachroum2020}.
 
 The majority of ICUs routinely collect data at relatively low temporal resolution, such as minute-by-minute. Therefore, mining this data to improve healthcare can have profound benefits \cite{Depreitere2016}. However, most artifact detection techniques for ICU data are based on high temporal resolution data (also referred to as waveform data). Signal Quality Index (SQI) based techniques rely on beat detection and template matching of periodic high-frequency waveforms \cite{Li2008,Zong2004}. Alternatively, statistical techniques such as robust Principal Component Analysis (rPCA) for univariate time series require periodic data \cite{Jin2017}. Therefore, these techniques do not apply to our low-temporal resolution non-periodic data. In \cite{Tsien2001}, artifact detection models are developed with decision trees based on both high and low-resolution data. The study shows that models built on low-resolution data can perform equally well on high-resolution data. However, the artifact detection task is formulated as a supervised learning problem, in which performance can be affected by the imbalanced distribution of artifactual vs non-artifactual samples.
 
 
We propose a hybrid artifact detection system for minute-by-minute resolution routinely collected ICU data, which consists of an LSTM-based variational autoencoder (VAE) and a statistical flatline detector. The statistical ``flatline'' component focuses on detecting flatline artifacts due to disconnections of sensors. We compare the performance of our system with two other systems based on autoencoder (AE) or ARIMA \cite{Zhou2018} models coupled with the flatline detection component. We also assess the impact of the threshold to detect artifacts and the regularization factor in the VAE results. We benchmark their performance on mean blood pressure signals (BPm) annotated by an expert.
 
 
 
 
\section{Materials and Methods}

\subsection{Dataset description}

We use a dataset from the KidsBrainIT project that was collected prospectively and extracted for research purposes from paediatric intensive care units (PICU). This is a multi-centre, multi-national dataset with patients from 16 PICUs in 7 countries \cite{Lo2018}. It includes 125 paediatric patients. From each patient, $\leqslant 14$ physiological signals were collected at minute-by-minute intervals  and artifacts marked by an expert researcher. The analysis presented here is applied to the BPm time-series of each selected record.

\subsection{Data preprocessing}
From the 125 patient files included in the original dataset, 85 are selected based on the availability of BPm recordings that contain at least $90\%$ of numerical data for the entire length of each recording. Consequently ``flatline'' errors are also considered as numerical data during this selection process as they cannot be distinguished just with their initial format.   The selection is made to exclude patient records that did not contain any recording of BPm or contained extensively large sections of missing BPm data. The selected records are then randomly split into 53 patient records for training, 15 patient records for validation and 17 patient records for testing.

For each patient's record separately, BPm is scaled by subtracting the median and then dividing by the Interquartile range (\nth{75} - \nth{25}). The scaling is applied at the record level to ensure that differences in individual blood pressure range are preserved when training the AE and VAE while reducing the effect of outliers. Each record in the training set is split into overlapping windows of $60$ minutes ($60$ samples in length $W$) with a sliding window step size of 1 minute.

For every record in the training set, the annotated labels are used to remove artifactual samples ensuring that the (V)AE-based component of the system is trained on clean patterns of physiological dynamics as opposed to artifactual ones. The windows of each record are shuffled and combined to ensure that the training is done in a patient agnostic manner. This results in a training matrix of windows with dimensions $N \times W$ where $W = 60$ and $N$ is the total number of windows. 


\subsection{Artifact Detection System}

\subsubsection{Statistical Component}

The statistical component has foundations in statistical anomaly detection \cite{Chandola2009}. Typically, a statistical method involves fitting a model to normal data and then using this model to test if new samples fit the model. In this study, a line is fitted on a unit step overlapping, sliding window of samples to calculate its gradient ($\nabla$), and then an artifactual label is attached to all samples in the window if $\mid \nabla \mid < 10^{-9}$  which indicates no fluctuation in the BPm time-series. The statistical component of the system is deployed directly on the validation and testing data without any interaction with training data.

\subsubsection{(V)AE-based Component}

 Long Short-Term Memory (LSTM) is a type of recurrent neural network that learns over long sequential data. An LSTM has, in addition to a hidden state, memory cells with their own recurrence network, which shares inputs and outputs with the main recurrent network and gates that enable efficient information flow \cite{Hochreiter1997}.

The (V)AE-based component consists of an LSTM-based encoder-decoder architecture. Training this component with artifact free data enables the architecture to learn a latent space which considers the temporal physiological dynamics of the signal, as the artefactual segments have been removed from the training set. A test sample is then mapped into this space and then reconstructed back. This sample is classified as an artifact if the reconstruction error ($\delta = \mid \mathbf{x} - \mathbf{x'} \mid$) is above a given threshold. This assumption has foundations in neural network-based and spectral anomaly detection techniques \cite{Chandola2009}.

An AE is a deep learning architecture for representation learning, utilizing the encoder and decoder format. The encoder maps a data point ($\mathbf{x}$) to its latent representation ($\mathbf{z}$), $\mathbf{z} = Encoder(\mathbf{x})$, and the decoder is a reverse mapping to the original data point, $\mathbf{x'} = Decoder(\mathbf{z})$ \cite{Goodfellow2016}. The complete approximation process is defined as: $\mathbf{x}~\simeq~Decoder(Encoder(\mathbf{x}))$.


An AE learns these mappings by minimizing a reconstruction loss between the actual input, $\mathbf{x}$, and its reconstruction, $\mathbf{x'}$.  A regularized AE can learn meaningful representations by adding a regularization term in the loss function \cite{Goodfellow2016}.

A VAE \cite{Kingma2014} is a form of regularized AEs that stems from probabilistic models, where a neural network approximates the joint distribution of observed ($\mathbf{x}$) and latent variables ($\mathbf{z}$), $p_\theta (\mathbf{x},\mathbf{z})$. The VAE learns the latent representation of the data by using an encoder, $q_\phi (\mathbf{z}|\mathbf{x})$, to approximate a Gaussian posterior distribution $p_\theta (\mathbf{z}|\mathbf{x})$. Sampling from this distribution, $\mathbf{z}$, and feeding them into the decoder, $p_\theta (\mathbf{x}|\mathbf{z})$, generates new meaningful samples, $\mathbf{x}$. The reparameterization step \cite{Kingma2014} enables the VAE to be trained by gradient backpropagation algorithm.  In this study we use the $\beta$-VAE variation \cite{Higgins2017} that utilizes a regularization coefficient $\beta$ and is trained by maximizing the Evidence Lower Bound (ELBO): 

\begin{equation*}
\label{ELBO_reg}
\mathcal{L}(\theta ,\phi ; \mathbf{x}) = \mathbb{E}_{q_\phi (\mathbf{z|x})}\left [ \log p_\theta (\mathbf{x|z}) \right ] - \beta D_{KL}(q_\phi (\mathbf{z|x})||p_\theta (\mathbf{z}))
\end{equation*}

Varying the parameter $\beta$ encourages the model to learn disentangled representations in the latent space with increasing $\beta$ leading to loss of higher frequencies \cite{Higgins2017}.

Here, we utilize the (V)AE-based component in different configurations of AE and VAE architectures. Across all tested configurations the dimensions of the encoder and decoder are kept the same in a symmetrical setup. The encoder and decoder consist of two LSTM layers, each with hidden dimension 64. The bottleneck dimension size is 12. These dimension sizes are tuned utilizing the validation set of patient records. 


\subsubsection{ARIMA Benchmarking}

To benchmark the operation of the (V)AE-based component to a state-of-the-art outlier detection technique, an ARIMA model \cite{Zhou2018,Blazquez-Garcia2021} is applied to the data and the respective performance metrics are reported. 

\subsubsection{Detection Merging}

The artifactual label outputs of the statistical and (VAE, AE, ARIMA) based components are finally merged using an OR logical function: A sample is labelled as artifact if it has been labelled as such by at least one of the components.

\subsection{Validation Set Tuning}

\subsubsection{Statistical Component Tuning}
The $15$ patient records in the validation set are used to tune the window size of this component. Window sizes of $5$, $10$, and $15$ samples are tested and the performance of this component is measured using the available annotations. Window size $10$ was selected for achieving the highest performance in the validation set.

\subsubsection{(V)AE Component Tuning}

The validation set is then used for the definition of a $\beta$ value range of the VAE configurations and a $\delta$ threshold percentile ($Q$) range for both the AE and VAE configurations based on which a sample will be labelled as valid or a ``spike'' error during the deployment of the (V)AE component. 

For $\beta$, the values $0.1$, $0.2$, $0.3$, $0.4$, $0.5$, $0.6$ are used after analysing the signal reconstructions for VAE configurations trained on the training records and deployed on the validation records. The starting point of $0.1$ is selected as a clear separation point from the operation of a traditional AE while the range stops at $0.6$ since the bandwidth of the reconstructed signal becomes significantly limited to low frequency fluctuations for $\beta > 0.6$.

For the definition of a $Q$ range, eight configurations (one for ARIMA, one for AE and six for VAE) are deployed on the validation set. A $\delta$ distribution is constructed per configuration and processed using the statistical component to remove the $\delta$ values of samples labelled as ``flatline'' errors. The operation of the statistical component is not expected to be perfect, however it removes the majority of ``flatline'' $\delta$ values to provide a $Q$ range that is calibrated for the detection of ``spike'' errors. Based on the resulting $\delta$ distributions, the $Q$ percentile range of $\nth{90}$, $\nth{92}$, $\nth{94}$, $\nth{96}$,$\nth{98}$  is selected, with each percentile consisting of an array of eight values that are used in the following testing set experiments, one for each configuration.

\subsection{Testing Set Experiments}

Using the parameter ranges defined with the validation set, the system is tested in a total of $40$ experimental setups based on all combinations of $\beta$ plus the vanilla AE, ARIMA and $Q$ values. Each experimental setup of (V)AE is repeated for five iterations and the performance of the system is measured through the calculation of the mean and standard deviation ($\sigma$) of the sensitivity and specificity metrics. The ARIMA model is applied to the data once. We consider artifactual samples the positive class and valid samples as the negative.

\section{Results and Discussion}

The results for the ARIMA, AE, and VAE with $\beta = 0.1$ models are displayed and compared in Table~\ref{ARIMA_AE_VAE}. Fig.~\ref{combined} displays the mean and $\sigma$ for both sensitivity and specificity calculated from the five iterations for each experimental setup of VAE $\beta$ and $Q$ values. The statistical component of the system remained the same across all experimental setups. 

\begin{table}
\caption{Mean ($Sensitivity | Specificity$) for ARIMA, AE and VAE $\beta = 0.1$}
\label{ARIMA_AE_VAE}
\setlength{\tabcolsep}{12pt}
\begin{tabular}{cccc}
\hline

Q & ARIMA & AE & VAE \\

\hline
$\nth{90}$ & $0.909 | 0.833$ & $0.942 | 0.835$ & $\textbf{0.950} | 0.854$  \\

$\nth{92}$ & $0.902 | 0.857$ & $0.937 | 0.857$ & $0.944 | 0.872$ \\

$\nth{94}$ & $0.887 | 0.882$  & $0.929 | 0.879$  & $0.933 | 0.889$  \\

$\nth{96}$ & $0.870 | 0.902$ & $0.919 | 0.899$ & $0.923 | 0.906$ \\

$\nth{98}$ & $0.845 | 0.924$ & $0.903 | 0.921$ & $0.905 | \textbf{0.926}$ \\

\hline

\end{tabular}
\end{table}

\subsection{Benchmarking VAE to AE and ARIMA}

\begin{figure}[!t]
\centerline{\includegraphics[width=\columnwidth]{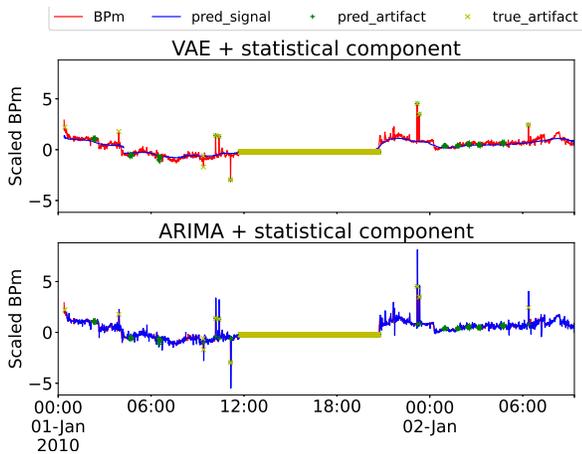}}
\caption{An example of BPm signal reconstruction and artifact detection}
\label{sample_signal_combined}
\end{figure}

When comparing the performance of the AE and ARIMA architectures to the VAE architecture with $\beta = 0.1$, the VAE  consistently outperforms the other architectures with regards to both its specificity and sensitivity with the performance increase being higher for specificity. In addition, the ARIMA-based model fits a model on each sliding window, making it very computationally heavy. Consequently, we would recommend the utilization of the VAE architecture over AE and ARIMA since it provides superior performance and increased control of its operation through the selection of the $\beta$ value. As in Fig.~\ref{sample_signal_combined}, the $\beta$ parameter is an attractive feature in detecting some types of artifacts, which tend to lie in the high frequency bands. However, a balanced $\beta$ value selection is required to avoid extensive low-pass filtering of the reconstruction. 

\subsection{VAE Configurations}

The $Q$ parameter controls how strict the threshold becomes for a sample to be labelled as an artifact based on its $\delta$. Higher values such as $Q = \nth{98}$ correspond to a less strict threshold as opposed to lower values such as $Q = \nth{90}$. As expected, increases in the value of $Q$ lead to higher values of specificity at the cost of reduced sensitivity. Out of the two parameters tested, $Q$ has the largest effect in the performance metrics.

Increasing the VAE $\beta$ leads to reconstructions with fewer high frequency components. Consequently, the specificity of the model increases due to a more clear definition between physiological fluctuations and artifactual ``spikes''. However a drop in sensitivity is also noted since certain ``spike'' errors can also be ``regularized'' in the reconstruction. The effect of $\beta$ is noticeably higher for lower $Q$ values $\nth{90}$, $\nth{92}$ with increased $\beta$ values favouring specificity at the cost of sensitivity as described. At $Q$ values of $\nth{96}$ and $\nth{98}$ changes in $\beta$ lead to minimal differences on specificity and sensitivity. 

\begin{figure}[!t]
\centerline{\includegraphics[width=\columnwidth]{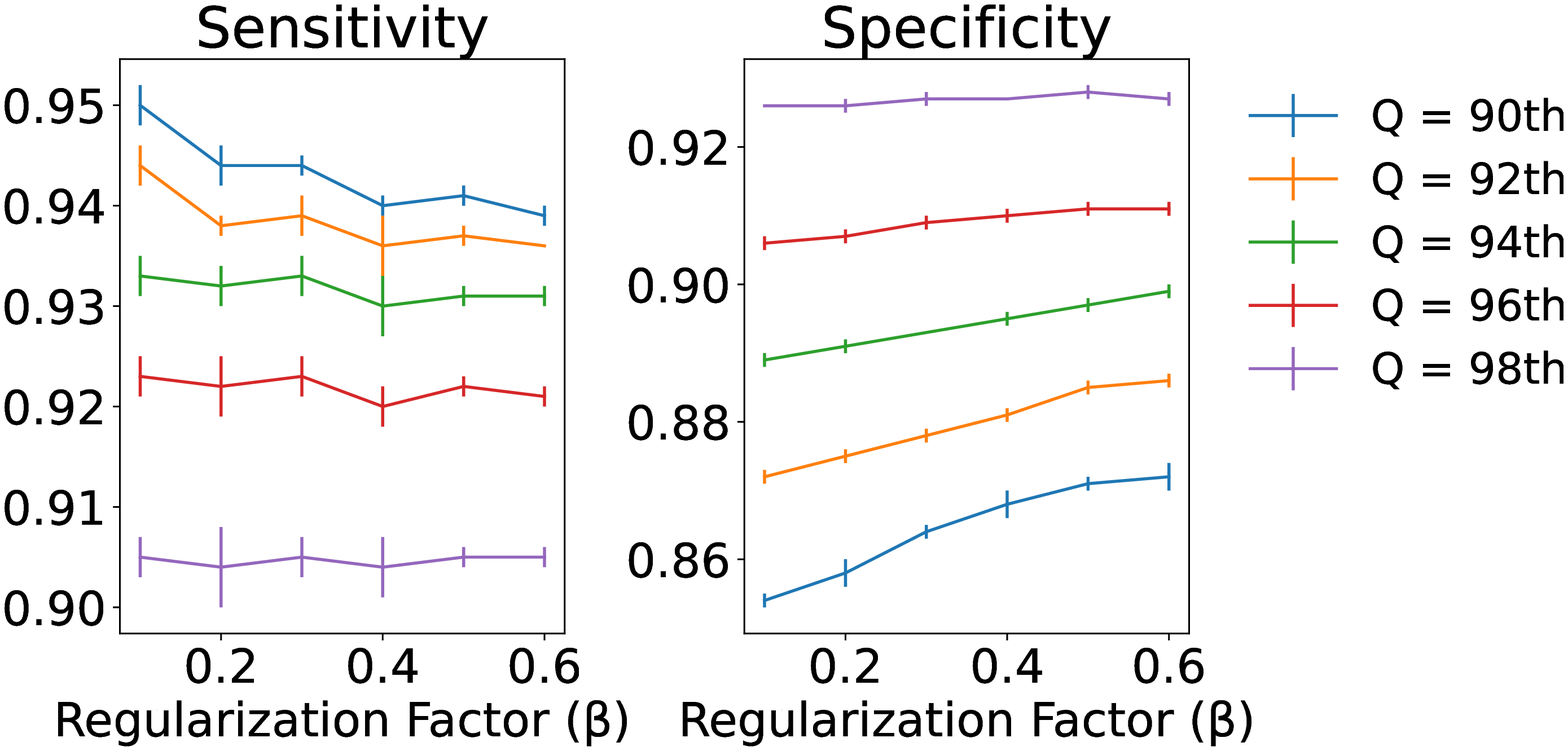}}
\caption{Mean sensitivity and specificity plotted for each combination of VAE: $Q$ and $\beta$ values. Points without error bars correspond to $\sigma < 10^{-3}$.}
\label{combined}
\end{figure}

Consequently, considering the effects of $Q$ and $\beta$ parameters and as shown in Fig.~\ref{combined} the highest value of specificity is achieved from the configurations with $Q = \nth{98}$ while the highest value of sensitivity is achieved with $Q = \nth{90}, \beta = 0.1$. Minor overlapping is observed in the values of sensitivity for configurations with $Q = \nth{92}$ and $\nth{94}$ at $\beta = 0.4$ across the five experimental iterations while no overlapping is observed for values of specificity. The maximum $\sigma$ observed across all experimental setups is $4 \cdot 10^{-3}$. Finally, the configurations with $Q = \nth{94}, \beta = 0.6$ and $Q = \nth{96}, \beta = 0.5$  achieve a balanced profile of specificity and sensitivity.

\subsection{Limitations of Study and Future Work}

The recorded performance in this preliminary study is the result of the combined output labels of the statistical and the VAE, AE or ARIMA based components of the hybrid detection system. Thus, the differences between the tested configurations can be dampened due to the steady performance of the statistical component that remains the same across all experimental setups. Due to the prevalence of ``flatline'' errors, the statistical component was important in ensuring that most of this kind of artefacts were detected.. For future studies we are interested in utilizing an architecture that could detect both ``flatline'' and ``spike'' errors using the same component to reduce the complexity of the system.

The presented system was implemented on the BPm time-series of the dataset. However, the available recordings include additional time-series formulated from Intracranial pressure (ICP), respiratory rate (RR), and Oxygen Saturation  signals. It would therefore be important to utilize the available time-series to verify the robustness of the approach across different physiological signals and to study the operation of a multivariate (V)AE-based component. 

Furthermore, the presented system is for offline operation. However, it provides a foundation for online applications in the future through the potential combination of saved offline pre-trained models with shorter online training batches.


\section{Conclusion}

This study reports preliminary results in the deployment of an artifact detection system consisting of a VAE-based and a statistical detection component for the detection of ``spike'' and ``flatline'' errors in BPm time-series. The presented system is successful in achieving offline detection of artifactual samples with both sensitivity and specificity $>90\%$. However, future research is required for the optimization of the system, to allow its utilization on datasets that do not contain training labels, and to extend it to multivariate time series.


\bibliographystyle{IEEEtran}
\bibliography{references}

\end{document}